\documentclass[amsmath,amssymb]{revtex4}

\usepackage[latin1]{inputenc}
\usepackage{graphicx}
\usepackage{dcolumn}
\usepackage{bm}
\usepackage{color}
 
\begin{document}

\title{Quantum Interrogation with particles.}
\author{Juan Carlos Garc\'ia-Escart\'in}
\email{juagar@tel.uva.es}
\author{Pedro Chamorro-Posada}
\affiliation{Departamento de Teor\'ia de la Se\~{n}al y Comunicaciones e Ingenier\'ia Telem\'atica. Universidad de Valladolid. Campus Miguel Delibes. Paseo del Cementerio s/n 47011 Valladolid. Spain.}
\date{\today}

\begin{abstract}
Interaction-free measurement and quantum interrogation schemes can help in the detection of particles without interacting with them in a classical sense. We present a density matrix study of a quantum interrogation system designed for particles that need not to be perfectly absorptive and compare the results to those of the usual setup.
\end{abstract}
\maketitle
\section{Introduction}
Measurement is a fundamental part of quantum physics. A particularly interesting phenomenon is that of interaction-free measurement, or IFM. With IFM we can obtain information on the state of an object without any interaction in the classical sense \cite{Dic81}. In \cite{EV93} Elitzur and Vaidman proved that IFM allows the detection of the presence of an object using a photon, even in the cases the photon does not interact with the object. In the Elitzur-Vaidman scheme the object, a bomb, is put in one of the arms of an interferometer. Depending on the presence or absence of the object the photon presents a different state at the output. 

Improved schemes for this ``Quantum Interrogation'' have been presented \cite{KWH95, KWM99, RG02}. The experiments have usually used macroscopic objects as the bomb \cite{KWH95, KWM99}, or high-finesse cavities \cite{TGK98} to guarantee the interaction of the light and the object. Here we will give a model to characterize a quantum interrogation scheme with particles that not always absorb passing photons, after the fashion of interrogation schemes for semitransparent objects \cite{MM01,Jan99,KSS00}. To do that density matrix formalism is used. 

This study will give the tools to carry out simpler quantum interrogation experiments. It is possible to recover the behaviour of the perfectly absorbing particle case at the cost of needing more cycles inside the interferometer. Section \ref{qint} analyzes the simple usual Quantum Interrogation system. Section \ref{partab} will generalize the results for partially absorptive bombs and give a recursive method for finding out the probabilities. The implications are briefly reviewed in the conclusion. 

\section{Quantum Interrogation}
\label{qint}
\begin{figure}[ht!]
\centering
\includegraphics{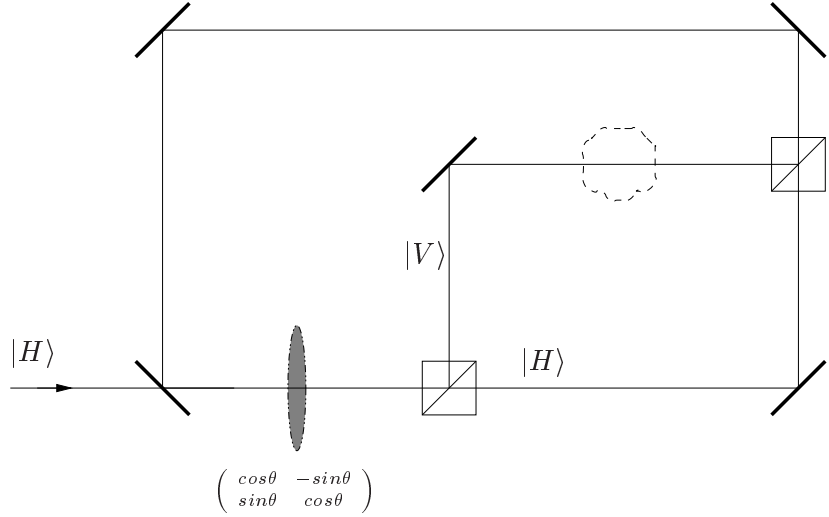}
\caption{Interaction free bomb detection system.}
\label{bomb}
\end{figure}

The interrogation system we will use is based on the proposal of \cite{KWM99}. Let's take a interferometer system like the one in figure \ref{bomb}. The particle (the bomb) can be present or not. 

We suppose that the original input is a horizontally polarized photon. The Hilbert space of the system will be given by the $\{|H\rangle,|V\rangle\}$ basis. In matrix representation
\begin{equation}
|H\rangle= \left( \begin{array}{c} 1\\ 0 \end{array} \right) , \hspace{2ex} |V\rangle= \left( \begin{array}{c} 0\\ 1 \end{array} \right).
\end{equation}

 The system has and two polarizing beamsplitters, PBS, that reflect vertically polarized photons while allowing the passage of horizontally polarized ones. The oval represents a polarization rotator, that gives the transformation $\vartheta= \left( \begin{array}{cc}
cos \theta & -sin \theta \\
sin \theta & cos \theta
\end{array} \right)$.

The interferometer has at its input the state $cos\theta|H\rangle+sin\theta|V\rangle$. If there is a bomb the probability of explosion is $sin^2\theta$, and the probability it doesn't explode is $cos^2 \theta$. In the latter case the state is reduced to $|H\rangle$ and the process starts again. If the photon undergoes $N$ cycles inside the interferometer, we have a probability of $cos^{2N}\theta$ of having $|H\rangle$ as the output state, with a total probability of explosion of $1-cos^{2N}\theta$. 

In the case with no bomb the superposition of states must be taken into account. If the photon goes through the circuit $N$ times the global effect will be given by the operator $\vartheta^N$. This operator can be obtained from the eigenvalues of $\vartheta$. 
\begin{equation}
\left| \begin{array}{cc}
cos \theta-\lambda & -sin \theta \\
sin \theta & cos \theta-\lambda
\end{array} \right|=(cos\theta-\lambda)^2+sin^2\theta=cos^2\theta-2cos\theta\lambda+\lambda^2+sin^2\theta=\lambda^2-2cos\theta\lambda+1=0
\end{equation}
From the characteristic equation $\lambda=\frac{2cos\theta\pm\sqrt{4cos^2\theta-4}}{2}=e^{\pm i\theta}$. The eigenvectors are $\frac{1}{\sqrt{2}}\left( \begin{array}{c} 1\\ i \end{array} \right)$ associated to $e^{-i\theta}$, and $\frac{1}{\sqrt{2}} \left( \begin{array}{c} 1\\ -i \end{array} \right)$ associated to $e^{i\theta}$. Therefore,
\begin{equation}
\vartheta=e^{-i\theta}\frac{1}{2}\left( \begin{array}{c} 1\\ i \end{array} \right)\left( \begin{array}{cc} 1& -i \end{array} \right)+e^{i\theta}\frac{1}{2}\left( \begin{array}{c} 1\\ -i \end{array} \right)\left( \begin{array}{cc} 1 &i \end{array} \right)=e^{-i\theta}\frac{1}{2}\left( \begin{array}{cc} 1&-i\\ i&1 \end{array} \right)+e^{i\theta}\frac{1}{2}\left( \begin{array}{cc} 1&i\\ -i&1 \end{array} \right),
\end{equation}
and, after $N$ cycles,
\begin{equation}
\vartheta^N=e^{-iN\theta}\frac{1}{2}\left( \begin{array}{c} 1\\ i \end{array} \right)\left( \begin{array}{cc} 1& -i \end{array} \right)+e^{iN\theta}\frac{1}{2}\left( \begin{array}{c} 1\\ -i \end{array} \right)\left( \begin{array}{cc} 1 &i \end{array} \right)=\left( \begin{array}{cc} \frac{e^{iN\theta}+e^{-iN\theta}}{2}& -\frac{e^{iN\theta}-e^{-iN\theta}}{2i}\\ \frac{e^{iN\theta}-e^{-iN\theta}}{2i} &  \frac{e^{iN\theta}+e^{-iN\theta}}{2}\end{array} \right)= \left( \begin{array}{cc} 
cos (N\theta) & -sin(N \theta) \\
sin (N\theta) & cos (N\theta)
\end{array} \right).
\end{equation}

The initial state $|H\rangle$ gives the state $cos(N\theta)|H\rangle+sin(N\theta)|V\rangle$ at the output. If $\theta=\frac{\pi}{2N}$, the output becomes $|V\rangle$. When $N\rightarrow \infty$ the probability of explosion (if there is a bomb present) tends to 0. For $N=24$ the probability of explosion is less than 10\%.

\section{Quantum Interrogation with partially absorptive particles}
\label{partab}
In this section we will address the quantum interrogation scheme of figure \ref{bomb} when the particle is a not perfectly absorbing one. In that case we will only have a probability $A$, smaller than one, of absorbing a passing photon. 

When there is no bomb the analysis of the previous section still holds. In the presence of a bomb, the situation is different from the perfect absorption case. Imagine we have a particle absorbing the photon with probability A. This absorption corresponds to bomb explosion. In this case, we can still have surviving $|V\rangle$ states, even if the particle is present. There is a probability 1-A of the photon not being absorbed.        

As our analysis will be based on the mixed state that appears after each round, we will study the evolution of the density matrix of the system. In order to be able to include all the cases we add the state $|B\rangle$, to represent that the atom has absorbed the photon. Now we have the basis $\{|H\rangle,|V\rangle,|B\rangle\}$. In matrix representation, 
\begin{equation}
|H\rangle= \left( \begin{array}{c} 1\\ 0\\0 \end{array} \right) , \hspace{2ex} |V\rangle= \left( \begin{array}{c} 0\\ 1\\0 \end{array} \right) , \hspace{2ex} |B\rangle= \left( \begin{array}{c} 0\\ 0\\1 \end{array} \right).
\end{equation}

It is possible to give the evolution of the probabilities using a recursive formula. For an arbitrary pure state $|\psi\rangle$ the density matrix is $\rho=|\psi\rangle\langle\psi|$. For a mixture of states $\rho=\sum_i p_i|\psi_i\rangle\langle\psi_i|$, where $p_i$ is the probability of having each of the states. The evolution under a unitary operator $U$ is given by $U\rho U^\dag$. If a measurement gives $m$ as a result, the new state will be $\frac{M_m\rho M_m^{\dag}}{tr(M_m^{\dag} M_m\rho)}$. This happens with probability $p(m)=tr(M_m^{\dag} M_m \rho)$. $M_m$ is the operator that projects the state to the subspace where the measurement is $m$, and $tr(\rho)$ denotes the trace of the density matrix.

We will consider the two distinguishable states of absorbing and not absorbing the photon. In a projective measurement $\sum_m M_m^{\dag}M_m=I$. In this case $M_B=|B\rangle\langle B|$, so
\begin{equation}
M_B= \left( \begin{array}{ccc} 0&0&0\\ 0&0&0\\0&0&1 \end{array} \right),\hspace{3ex} M_{\bar{B}}= \left( \begin{array}{ccc} 1&0&0\\ 0&1&0\\0&0&0 \end{array} \right).
\end{equation}

We will assume that, after each round, we take a projective measurement of the subspaces given by these two matrices. This is equivalent to having a continuous measurement on the absorbed state (for instance fluorescence observation of the excited level). We concentrate on this model because it gives a better picture for the cases in which the particle exhibits a quantum behaviour and is in a superposition of being and not being in the arm of the interferometer. An alternative model for semiabsorptive particles is considering a particle that with probability A acts as a measurement, like in the example of section \ref{qint}, and has no effect with probability 1-A. In that case, the density matrix analysis can be applied with slight modifications, as well as Markov chain techniques. It is interesting to note that, although there are small variations in the behaviour for a given N, the asymptotic results for large N are maintained. 
 
There are two different evolutions. If we detect the explosion the procedure is halted, so the state remains $|B\rangle$. The evolution is given by the identity matrix, I. In the no-absorption subspace the evolution is given by two terms. The first is the $U$ operator (here the polarization rotator's evolution plus a term that tells that once we are in the absorption state we cannot change the state).
\begin{equation}
U= \left( \begin{array}{ccc}
cos \theta & -sin \theta &0\\
sin \theta & cos \theta &0\\
0&0&1
\end{array} \right).
\end{equation}
Then we add the absorption matrix 
\begin{equation}
Ab= \left( \begin{array}{ccc}
1 & 0 &0\\
0 & \sqrt{1-A}  & -\sqrt{A}\\
0&\sqrt{A}&\sqrt{1-A}
\end{array} \right).
\end{equation}
Notice that, because of the unitarity condition, Ab includes some emission terms where $|B\rangle$ becomes $-|V\rangle$. It is the continuous measurement process that ensures that the transition to $|B\rangle$ is irreversible.  

The system evolves according to $\rho_{i+1}=M_B\rho_i M_B^{\dag} + Ab U M_{\bar{B}} \rho_i M_{\bar{B}}^{\dag} U^{\dag} Ab^{\dag}$, indicating the evolution from round i to round i+1.

After $N$ cycles, the probability of finding a photon in $|H\rangle$ is $tr(M_H M_H^{\dag}\rho_N)$, and we find it in $|V\rangle$ with probability $tr(M_V M_V^{\dag}\rho_N)$. There is a probability $tr(M_B M_B^{\dag}\rho_N)$ of it having been absorbed during the process. We can simulate the evolution from an initial state $|H\rangle$ ($\rho_0=\rho_H$).

\begin{figure}[ht!]
\centering
\includegraphics[height=10cm]{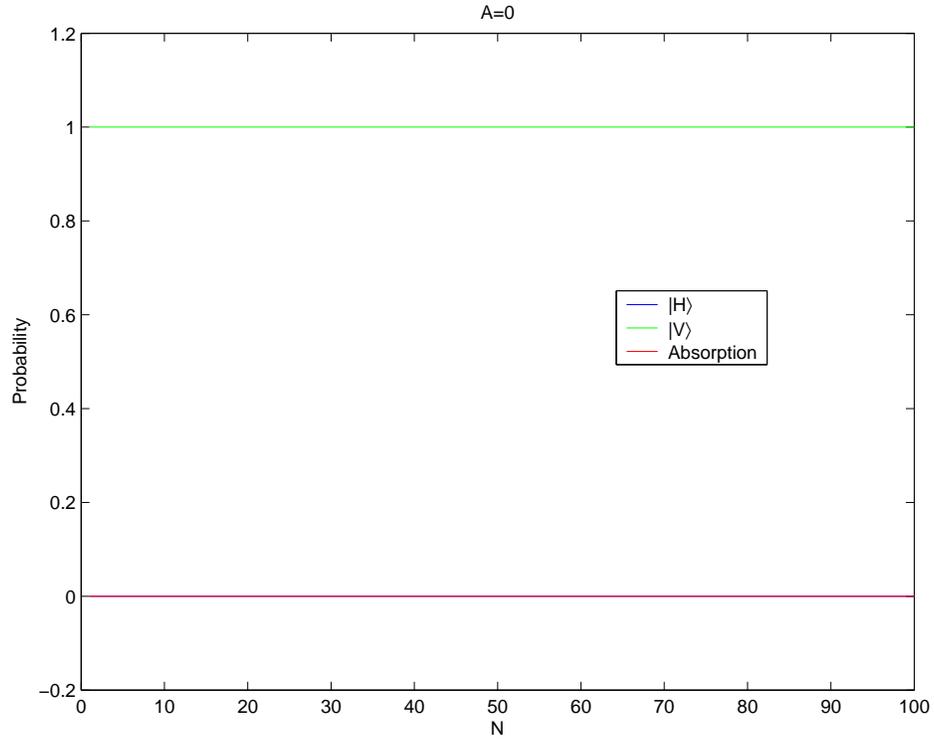}
\caption{Probability of the different output states for an atom with an interaction probability of 0 as a function of the number of cycles.}
\label{A0}
\end{figure}

\begin{figure}[ht!]
\centering
\includegraphics[height=10cm]{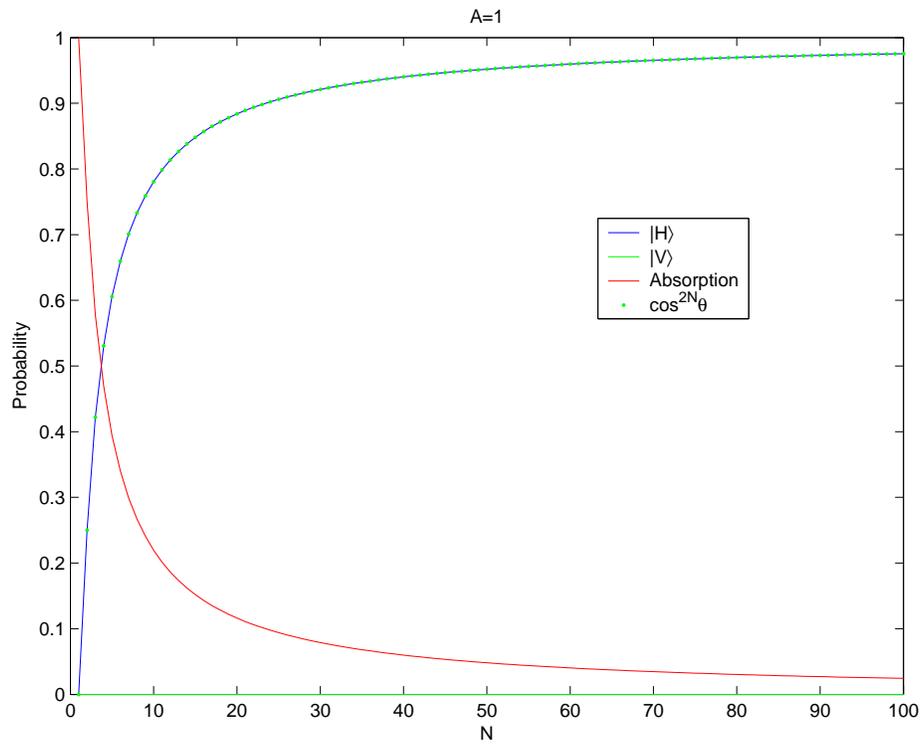}
\caption{Probability of the different output states for an atom with an interaction probability of 1 as a function of the number of cycles.}
\label{A1}
\end{figure}

Figure \ref{A0} shows the results for $A=0$ (equivalent to the case with no particle). The results concur with those of our previous analysis. When $A=1$ (figure \ref{A1}) we have the same result as in the perfectly absorptive particle case.\\

\begin{figure}[ht!]
\centering
\includegraphics[height=10cm]{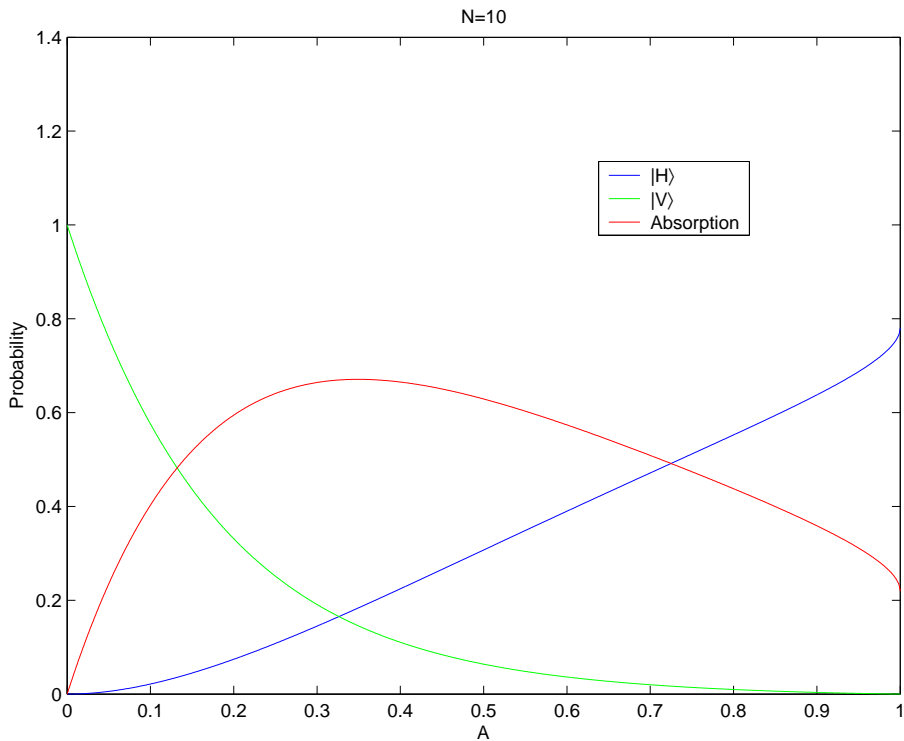}
\caption{Probabilities for different values of $A$ after 10 cycles.}
\label{N10}
\end{figure}

\begin{figure}[ht!]
\centering
\includegraphics[height=10cm]{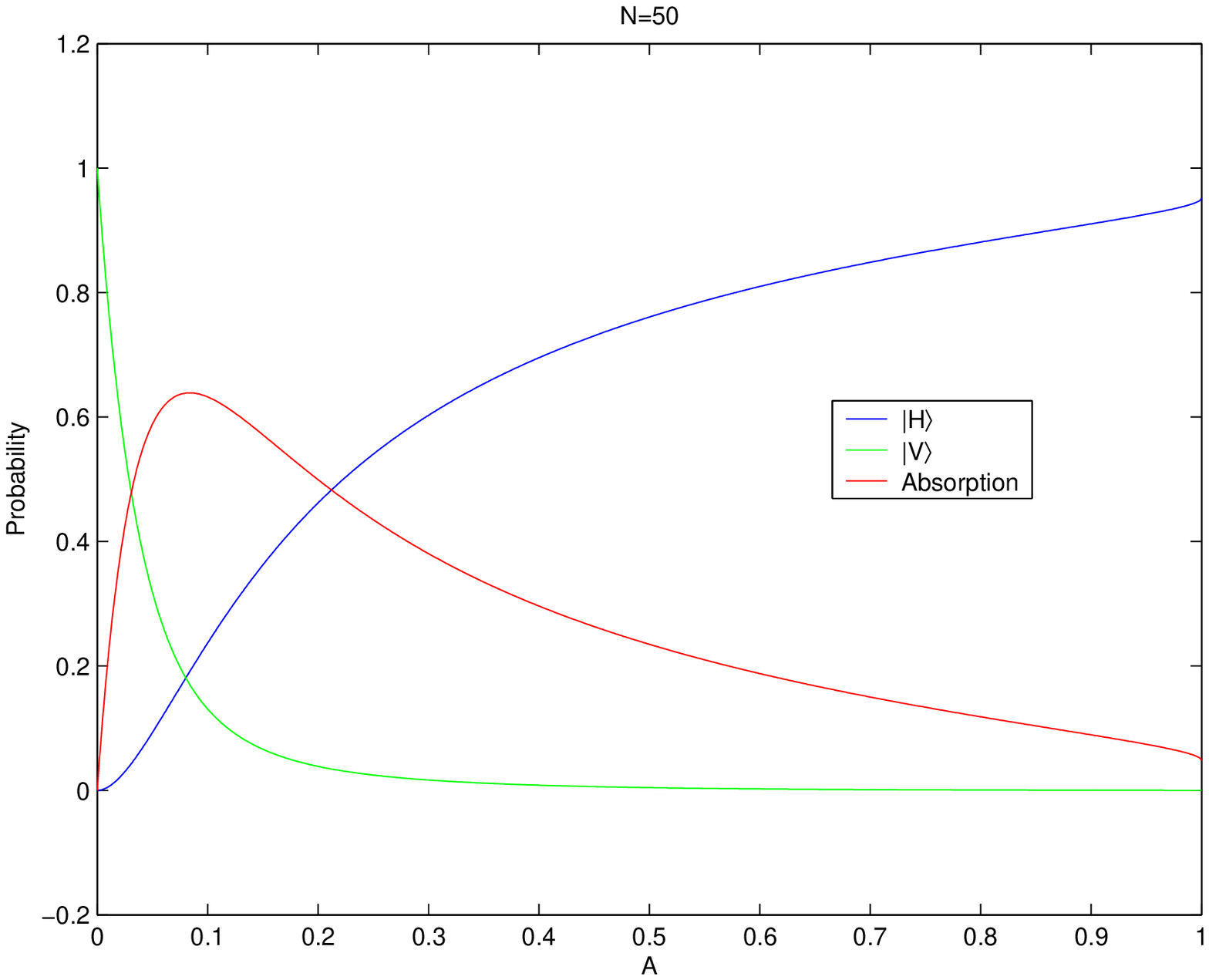}
\caption{Probabilities for different values of $A$ after 50 cycles.}
\label{N50}
\end{figure}

\begin{figure}[ht!]
\centering
\includegraphics[height=10cm]{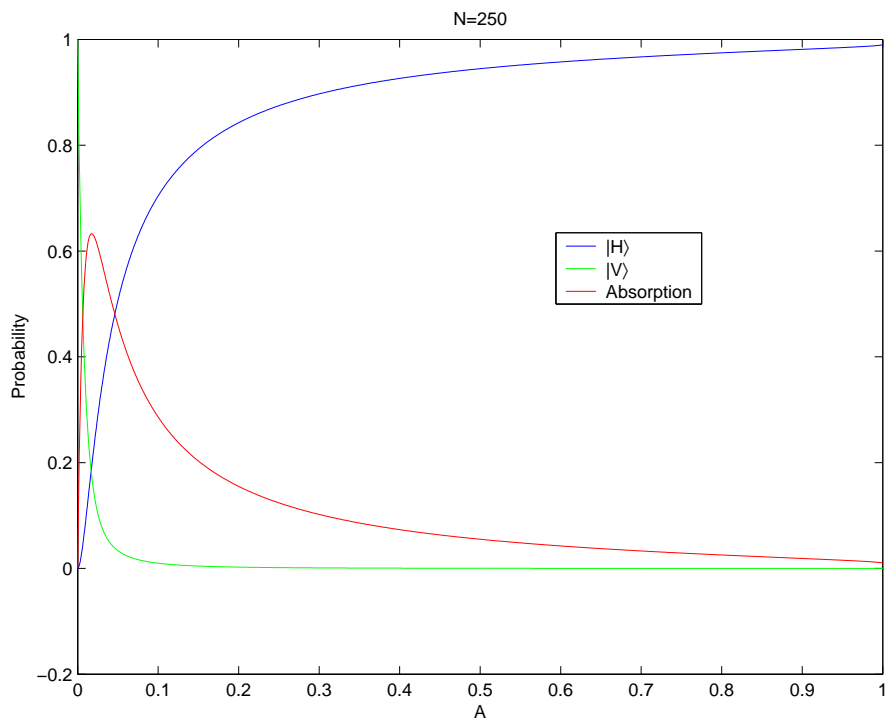}
\caption{Probabilities for different values of $A$ after 250 cycles.}
\label{N250}
\end{figure}

Figures \ref{N10}, \ref{N50} and \ref{N250} show how different values of $A$ affect to the value of the probabilities of the output states, for a different number of cycles.\\

\begin{figure}[ht!]
\centering
\includegraphics{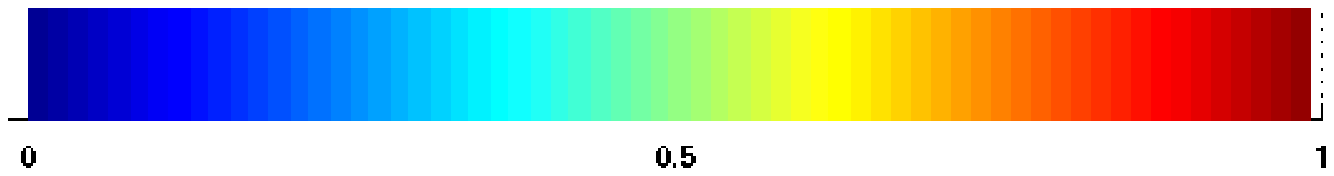}
\includegraphics[height=10cm]{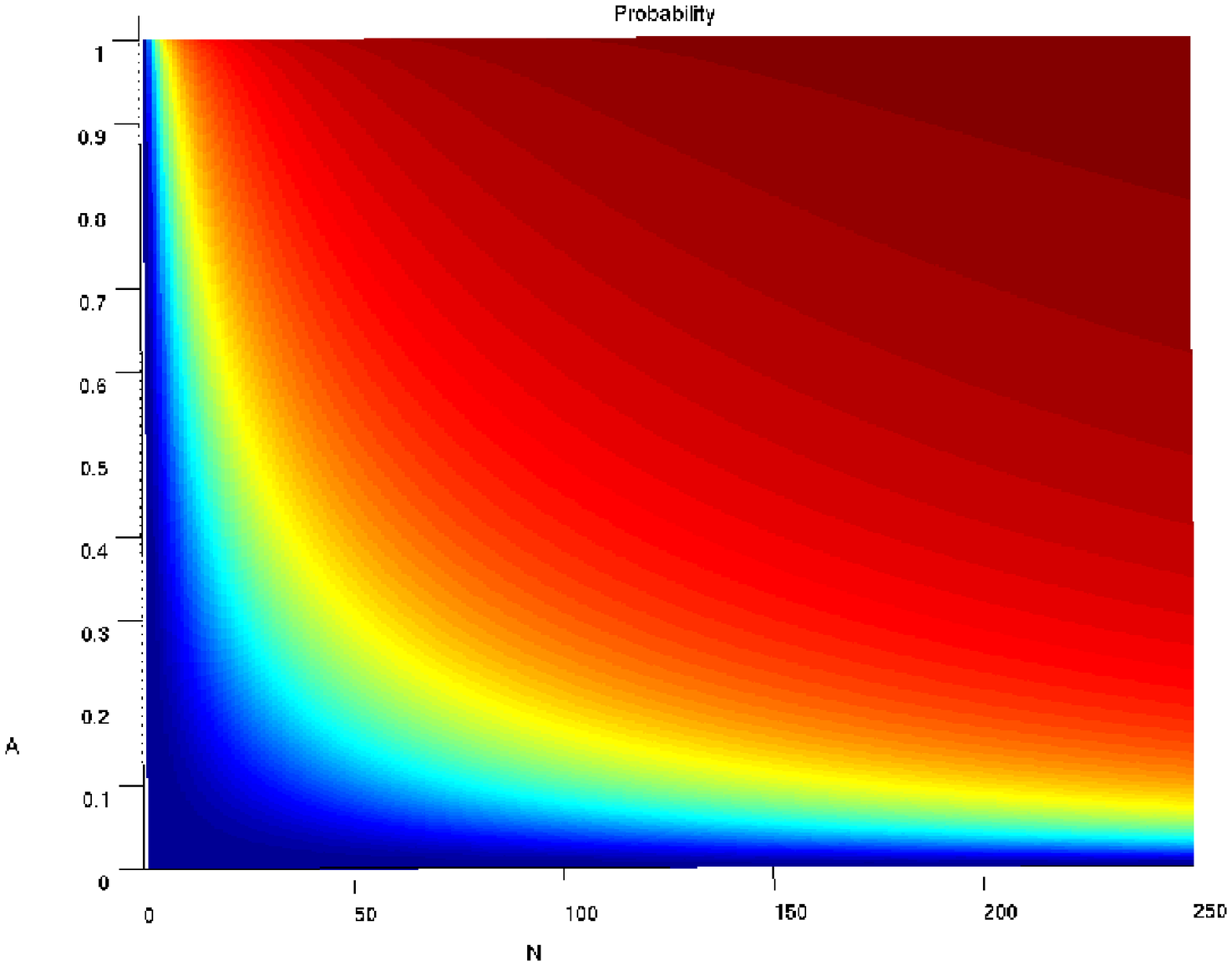}
\caption{Probability of finding the photon in $|H\rangle$ when there is a particle in the upper arm as a function of $A$ and $N$. Warmer colors correspond to higher probability values}
\label{nivel}
\end{figure}

We can see that the higher the absorption probability $A$, the better the performance of the system, but, as we have a prediction tool, we can fix the number of cycles we need depending on the desired probability values and the absorption probability of our particle. The growth of the probability towards the ideal case (with no absorption, and the photon going out in different polarizations when there is a particle and when there is none) can be seen in figure \ref{nivel}. There, we can see with a color code the evolution for different values of $A$ and $N$.

\section{Conclusion}
We have seem that quantum interrogation schemes can use particles that are not totally absorptive. Those particles could be atoms, electrons in a potential well, or small molecules. Any system able to absorb light with a certain, greater than 0, probability can be used as the bomb. The evolution of the system can be described using density matrix formalism, and a recursive formula for the state has been derived. This opens the door to simpler experiments where the bomb is a quantum object able to show superposition, a case that has been theoretically analysed \cite{ZZF01,Har92} but with few experimental results.
\newpage
\bibliographystyle{apsrev}
\bibliography{lspace} 

\end{document}